\documentclass[aps,twocolumn,nofootinbib,superscriptaddress]{revtex4-1}

\usepackage{amsmath,amsfonts,amssymb}
\usepackage{wrapfig}
\usepackage{graphicx}
\usepackage{bbm}
\usepackage{graphics}

\usepackage{braket}

\begin{document}

\title{Comment on ``Using an atom interferometer to infer gravitational entanglement generation''}
\author{Daniel Carney}
\affiliation{Physics Division, Lawrence Berkeley National Lab, Berkeley, CA}
\author{Holger M\"uller}
\affiliation{Department of Physics, University of California, Berkeley, CA}
\author{Jacob M. Taylor}
\affiliation{Joint Quantum Institute, NIST, Gaithersburg, MD}
\date{\today}

\begin{abstract}
Our paper \cite{carney2021using} contains a technical error which changes some of the conclusions. We thank Streltsov, Pedernales, and Plenio for bringing the essence of this error to our attention. Here we explain the error, examine its consequences, and suggest methods to overcome the resulting weakness in the proposed experiment. 
\end{abstract}

\maketitle

\textbf{Introduction.}
In brief, it is possible to construct a channel on the qubit-oscillator system which satisfies all of our requirements (time-translation invariance, preservation of $\sigma_z$, and separable), which nevertheless exhibits (partial) collapse-and-revival dynamics. Thus observation of a collapse-and-revival signal is naively insufficient to conclude that the underlying channel is capable of entanglement generation. For clarity, we give an explicit construction of such a channel here. The fact that this counter-example can be constructed is a symptom of a technical error in the proof of theorem 1 of \cite{carney2021using}.

The correct version of the theorem is detailed below; one has to make a stronger assumption about the specific form of the Lindblad equation. This weaker theorem is still strong enough to rule out interesting cases, such as models where gravity arises as a stochastic force (we provide a detailed example below). However, it does allow for counterexamples. The counterexample constructed here (based on LOCC channels \cite{kafri2014classical}) requires adding additional noise sources into the model, which will partially destroy the revival signature. Thus, seeing revival above this loss would still rule out such a non-entangling model; our proposed experiment is still an optimal path to this goal. Thus the experiment may still be capable of ruling out this counter-example class of models, but further work will be needed to ensure both that this class is the only possible type of counter-example, and that all models of this type are excludable through the revival signature.

\textbf{Formal issue.} The central technical result of the paper is stated as a theorem about two-body channels $L$ obeying a certain separability criterion. This was assumption (c) in the paper, viz. 
\begin{quote}
(c) $L$ is a separable channel: all of its Krauss operators are simple products. In particular, this means that any initial separable (non-entangled) state evolves to a separable state: $\rho(t) = L_t[\rho(0)]$ is separable for all separable initial states $\rho(0)$.
\end{quote}
Given such a separable Krauss representation exists, say $L[\rho] = \sum_i L_i \rho L_i^\dag$ with $L_i = A_i \otimes B_i$, it is clear that the channel is non-entangling in the sense that it maps separable density matrices to separable density matrices. We then imposed a time-translation invariance assumption to write the time evolution under this map in Lindblad form. To do this, one expands the Krauss operators to lowest order in $dt$ as
\begin{equation}
\label{expansion}
L_0 = 1 - i H dt + K dt, \ \ \ L_i = E_i \sqrt{dt}, \ \ \ K = \frac{1}{2} \sum_{i \geq 1} L_i^\dag L_i.
\end{equation}
We then applied the Krauss separability condition $L_i = A_i \otimes B_i$ to this expression to derive constraints on the Lindblad operators $E_i$.

The essential issue is that the separability criterion (c) need only be true in one particularly basis of Krauss operators, which is not generally the format given by the Kraus operators $L_j$ above. Any two sets of Krauss operators $\{L_i\}, \{L'_j\}$ related by unitary transformations $K_i = \sum_{j} U_{ij} L'_j$ produce equivalent channels. These unitaries can be non-local, and thus a channel with a separable representation can be unitarily equivalent to a Lindblad form that does not satisfy separability in $H$ and $K$ in $L_0$. The problem is that computing the Lindblad representation of a channel \eqref{expansion} will generally require a basis rotation to put one of the Krauss operators $L_0$ into the specific form $L_0 = 1 + \mathcal{O}(dt)$. In particular, a given non-separable Krauss representation may require a non-local unitary rotation before admitting the expansion \eqref{expansion}, and thus the constraints derived on the Lindblad operators in the paper are not generally correct.

Thus, we correct the theorem, weakening it significantly, replacing assumption (c) above with a more explicit assumption (c'), expressed directly at the level of the Lindblad equation:
\begin{quote}
(c') $L$ is generated by a separable Lindblad evolution: the differential time evolution $\rho(t) = L[\rho(t_0)]$ is given by
\begin{equation}
\label{lindblad}
\dot{\rho} = -i [H,\rho] + \sum_i E_i \rho E_i^\dag - \frac{1}{2} \{ E_i^\dag E_i, \rho \}
\end{equation}
with the Hamiltonian $H$ and Lindblad operators $E_i$ all taking the form of product operators and $\sum_i E_i^\dag E_i$ is a sum of operators acting on only $A$ or $B$, not both.
\end{quote}
Everything in the rest of the paper would then follow without change. (We also note that using such a Lindblad form already assumed the time-translation semigroup assumption (a) used in the paper, which would then be redundant). In particular, any such channel necessarily leads to a monotonically decaying atomic visibility. 

This weakens, but does not trivialize, the basic result. It is possible to make a model which reproduces semi-classical gravitational interactions which takes the Lindblad form \eqref{lindblad}, and observation of the collapse and revival would rule this out. To see a concrete example, consider the experiment discussed in the paper, with a mechanical oscillator gravitationally coupled to an atom localized to one of two locations. We can construct a Lindbladian model with a purely stochastic gravitational force, that is, the force arises due to random application of impulses whose direction depends on . Let $E_{0} = e^{- i \alpha x} \ket{0} \bra{0}$ and $E_{1} = e^{+ i \alpha x} \ket{1} \bra{1}$, with $\ket{0,1}$ denoting the two atom locations and $\alpha$ a constant to be fixed shortly. These are product operators and thus produce a separable Lindblad equation. The evolution of the mechanical momentum under damping from $\{E_0,E_1\}$ with rate $\gamma$ is
\begin{equation}
\dot{p} = \gamma \sum_{i} E_{i}^\dag p E_{i} - \frac{1}{2} \{ E_{i}^\dag E_{i}, p \} = \gamma \alpha \sigma_z,
\end{equation}
so setting the coefficient $\alpha = G_N m M \ell x_0/d^3 \gamma$ produces the expected Newtonian force on the pendulum, dependent on the atomic position.\footnote{Similarly, a stochastic gravitational force felt by the atoms can be modeled by including Lindblad operators like $E(z) = P(z) e^{i \beta x \sigma_z}$, where $P(z) \sim e^{-(x-z)^2/x_0^2}$ is a projector that localizes the mechanical position to a window near $z$.} In particular, this would give a ``semiclassical'' evolution like $\braket{\dot{p}} = \gamma \alpha \braket{\sigma_z}$. This evolution is incapable of generating revival dynamics [consistent with the theorem assuming (c')], and thus can be directly ruled out experimentally in the manner described in the paper.

However, the (correct) Lindblad version of the theorem is a strictly weaker result: it does allow for construction of channels which are non-entangling yet still capable of producing the revival signals. We now give an explicit example of this. The example also suggests a way to close this loophole: such a channel requires adding certain extra noise terms to the evolution. The atomic interferometry experiment we originally proposed would be also capable of ruling these terms out, because these noise terms cause a partial loss of the revival signal.

\textbf{Explicit example.} We now construct a manifestly non-entangling channel which obeys the time-translation semigroup law, preserves $\sigma_z$, and still allows for partial collapse-and-revival dynamics. This is a measurement-and-feedback channel which uses only local operations. Concretely, consider two channels, one with Krauss operators
\begin{equation}
K_{\pm} = e^{\pm i \beta \sigma_z \sqrt{dt}} [\cos (\alpha x \sqrt{dt}) \pm \sin (\alpha x \sqrt{dt})]/2,
\end{equation}
and the other with
\begin{equation}
K'_{\pm} = e^{\pm i \alpha x \sqrt{dt}} [\cos (\beta \sigma_z \sqrt{dt}) \pm \sin (\beta \sigma_z \sqrt{dt})]/2.
\end{equation}
The coefficients $\alpha,\beta$ will be fixed shortly. These are both valid channels $\sum_{a} K_a^\dag K_a = 1$,  $\sum_a K_a^{'\dag} K'_a = 1$, and are manifestly separable. These arise by Ramsey-type measurements on either the qubit or oscillator, with outcomes $\pm 1$, followed by feedback onto the other system. Concatenation of these channels thus forms a new channel $L$ with four Krauss operators $L_{ab}$, namely
\begin{equation}
L_{ab} = K'_a K_b.
\end{equation}
These concatenated Krauss operators are again manifestly separable. Moreover, the channel clearly preserves $\sigma_z$, since the Krauss operators commute with $\sigma_z$. We just need to check if this channel generates Lindblad-form evolution. Notice that none of these Krauss operators have an expansion like that used for $L_0$ in \eqref{expansion}. However, direct computation gives
\begin{align}
\begin{split}
d\rho = & L[\rho(t_0)] - \rho(t_0) \\
 = & -2 i \alpha \beta [x \sigma_z, \rho] dt - \alpha^2 ( 2 x \rho x - \{ x^2,\rho \}) dt \\
& - \beta^2 ( 2 \rho - 2 \sigma_z \rho \sigma_z ) dt + \mathcal{O}(dt^{3/2}),
\end{split}
\end{align}
which we recognize as a Lindblad equation \eqref{lindblad}, with
\begin{equation}
\label{example}
H = -2 \alpha \beta x \sigma_z, \ \ \ E_1 = i \sqrt{2} \alpha x, \ \ \ E_2 = \sqrt{2} \beta \sigma_z.
\end{equation}
We see that while the Lindblad operators $E_i$ are separable, the Hamiltonian term is not. This is despite the fact that, by construction, this channel is incapable of generating qubit-oscillator entanglement, since it admits a separable Krauss representation.

The Hamiltonian term \eqref{example} is precisely the one used in our paper to generate collapse-and-revival signals. See equation (2). Thus this non-entangling channel can at least partially mimic the signal expected from the full quantum channel where we have only the Hamiltonian term.

\textbf{Detection of additional noise.} However, we see that this new non-entangling channel includes noise operators, and the strength of this noise is fixed by the same coupling constant as the Hamiltonian interaction. Concretely, comparing to the realization of the pure Hamiltonian with an atom and mechanical oscillator as proposed in our paper, the product of $\alpha$ and $\beta$ is fixed to
\begin{equation}
\alpha \beta = \frac{G_N m M \ell}{d^3},
\end{equation}
see equation (23) [note that here we are using the usual dimensionful position operator $x = x_0 (a + a^\dag)$]. On dimensional grounds, noting that $M_{\rm pl}^2 = 1/8\pi G_N$, a natural splitting of the coefficients would be
\begin{equation}
\label{couplings}
\alpha = \frac{M}{\sqrt{8\pi} M_{\rm pl}} \frac{1}{d^{3/2}}, \ \ \ \beta = \frac{m}{\sqrt{8\pi} M_{\rm pl}} \frac{\ell}{d^{3/2}}.
\end{equation}
In principle, other choices are possible, although they would seem to correspond to non-local measurements. Note that we cannot set either $\alpha$ or $\beta$ to zero without turning off the induced gravitational coupling, and thus we inevitably get the addition of noise on both the atom and oscillator.

In particular, the additional noise will cause a degradation of the revival signature. Because of the Lindblad operator $E_1$, the oscillator will experience anomalous heating. With the specific choice \eqref{couplings}, the decay constant from this heating is
\begin{equation}
\label{heating}
\gamma = \alpha^2 x_0^2 = 3 \times 10^{-5}~{\rm Hz} \times \left( \frac{M}{1~{\rm mg}} \right) \left( \frac{1~{\rm mHz}}{\omega} \right) \left( \frac{1~{\rm mm}}{d} \right)^3.
\end{equation}
In the ``unboosted'' protocol, this heating would have to be gravitationally transmitted to the atom, leading to a loss of coherence of order $\lambda^2 \gamma \sim G_N^3$, which is negligibly small. There would also be a direct anomalous heating of order $G_N^2$ acting on the atoms, which again is negligibly small.

However, in the ``boosted'' protocol, we would have an anomalous heating rate \emph{linear} in $G_N$, and it appears that this could be detected in the experiment we have proposed. To see this, recall that the unboosted protocol begins by using a non-gravitational coupling to produce an entangled state of the form
\begin{equation}
\ket{\psi} = (\ket{\delta, 0} + \ket{-\delta, 1})/\sqrt{2},
\end{equation}
where $\ket{0,1}$ are the two atom locations and $\ket{\pm \delta}$ are displaced coherent states of the oscillator [see equation (7), with $g \to g'$ some controllable non-gravitational coupling]. To see what this will do to the atomic visibility, we can write the initial total density matrix in the qubit basis as
\begin{equation}
\rho(0) = \frac{1}{2} \begin{pmatrix} \ket{\delta} \bra{\delta} & \ket{\delta} \bra{-\delta} \\ \ket{-\delta} \bra{\delta} & \ket{-\delta} \bra{-\delta} \end{pmatrix}.
\end{equation}
Neglecting any other evolution, the oscillator heating \eqref{heating} will cause dephasing of the oscillator, leading to a state of the form
\begin{equation}
\rho(t) = \frac{1}{2} \begin{pmatrix} \ket{\delta} \bra{\delta} & e^{-\gamma t/2} \ket{\delta} \bra{-\delta} \\ e^{-\gamma t/2} \ket{-\delta} \bra{\delta} & \ket{-\delta} \bra{-\delta} \end{pmatrix}.
\end{equation}
This oscillator dephasing thus causes a loss in the atomic visibility with a decay rate $\gamma \sim G_N$. With $N$ atoms, and a locally-acting gravitational channel, we have $N$ Krauss operators $K_{\pm} \to \{ K^i_{\pm} \}_{i=1,\ldots,N}$ and similarly for $K'_{\pm}$. This leads to an effective heating rate $N \gamma$. So for example with $N = 10^8$ atoms and a more realistic choice of a $1~{\rm Hz}$ oscillator, we are looking for anomalous heating with a rate of order $1~{\rm Hz}$. For a reasonably high-$Q$ mechanical oscillator, this should be clearly visible above ordinary thermal noise. In words, the ``boosted'' experiment performs an entanglement-assisted measurement of anomalous heating (in this case, gravitational anomalous heating).

\textbf{Summary and conclusions.} The original claim in our paper was that observation of collapse-and-revival dynamics in the atomic interferometry proposed there experiment would rule out \emph{any} model in which gravity does not generate entanglement. This conclusion was too strong; as explained above, the correct statement is that a revival only directly rules out any model with separable Lindblad operators. 

Although this does include a large class of models in which gravity is purely stochastic, it leaves open possible counter-examples constructed using LOCC measurement-and-feedback channels. However, we have seen that in explicit such LOCC channels, additional anomalous gravitational noise sources tend to be present. The experiment proposed in our paper is well-optimized to rule out such extra noise, because it would generate a loss of atomic revival at a rate linear in $G_N$.

We thus close with a key open question: are all non-entangling channels consistent with Lorentz invariance and the (semiclassical) Einstein field equations always of this LOCC form \cite{kafri2014classical}? If so, the proposed experiment would still be capable of conclusively demonstrating that the gravitational interaction is entangling, because it could rule out the only possible alternative consistent with the description of the two-body interactions as a quantum channel. This is a deep question well beyond the scope of this errata, and we leave it to future work.

\bibliography{gravity-comment.bib}

\end{document}